\documentclass{article}
\usepackage{amssymb}
\usepackage{amsmath}
\usepackage{amsfonts}
\usepackage{authblk}
\newtheorem{proposition}{Proposition}
\def\BOne{{\mathchoice {\rm 1\mskip-4mu l} {\rm 1\mskip-4mu l}
                          {\rm 1\mskip-4.5mu l} {\rm 1\mskip-5mu l}}}
\begin{document}
\title{Sasaki-Ricci flow equation on five-dimensional Sasaki-Einstein space $Y^{p,q}$}

\author{Mihai Visinescu\thanks{mvisin@theory.nipne.ro}}

\affil{Department of Theoretical Physics,

National Institute for Physics and Nuclear Engineering,

Magurele, P.O.Box M.G.-6, Romania}

\date{}
\maketitle

\begin{abstract}

We analyze the transverse K\"{a}hler-Ricci flow equation on Sasaki-Ein\-stein space 
$Y^{p,q}$. Explicit solutions are produced representing new five-dimensional
Sasaki structures. Solutions which do not modify the transverse metric preserve the 
Sasaki-Einstein feature of the contact structure. If the transverse metric is altered,
the deformed metrics remain Sasaki, but not Einstein.

\end{abstract}

~





\section{Introduction}

The concept of Ricci flow was originally introduced by Hamilton in 1982 \cite{RH} representing
a major tool that allows to continuously deform a Riemannian manifold.

The complex analogue of Hamilton's Ricci flow, known as K\"{a}hler-Ricci flow was introduced by Cao 
\cite{CAO} to give a parabolic proof of the Calabi-Yau theorem. Eventually the Ricci flow, as a 
method to deform Riemannian metrics, was applied to Sasaki manifolds \cite{SWZ,FOW}.

In physics this tool has gained significant attention in the framework of general relativity,
renormalization group equations, evolution of wormholes and black holes, cosmological models, etc.

In the last time Sasaki-Einstein geometry, as an odd-dimensional cousin of K\"{a}hler-Einstein geometry,
has played an important role in the AdS/CFT correspondence. There has been a growing interest in
contact geometry to study mechanical systems when the Hamilton function explicitly depends on time.
Contact geometry is also used in thermodynamics, in the description of systems with dissipation, 
geometric optics, geometric quantization, control theory, etc (see e.g. \cite{AB} for a recent review
of applications of contact Hamiltonian dynamics in various fields).

In the present paper we study the transverse K\"{a}hler-Ricci flow on Sasaki-Einstein space $Y^{p,q}$.
In the framework of AdS/CFT correspondence, $Y^{p,q}$ spaces have been employed to
provide an infinite class of dualities. In order to investigate the Sasaki-Ricci flow equation 
we introduce a set of local complex coordinate to parametrize the transverse holomorphic structure
and the Sasakian analogue of the K\"{a}hler potential for the K\"{a}hler geometry.

In spite of the complexity of the K\"{a}hler-Ricci flow equation, we find some particular explicit
solutions. If the transverse part of the metric is not changed, the deformed $Y^{p,q}$ remains
Sasaki-Einstein. In the opposite case, a modification of the transverse part of the metric leads
to a Sasakian structure, but not Einstein.

The paper is organized as follows. In the next Section we recall the definitions and main facts 
about Sasakian structures and Sasaki-Ricci flow. In Section 3 we investigate the transverse 
K\"{a}hler-Ricci flow equation on the Sasaki-Einstein space $Y^{p,q}$ and produce families of 
deformed metrics. The paper ends with conclusions in the last Section.

\section{Background}

In this section we recall the key concepts in the theory of Sasaki manifolds and transverse
K\"{a}hler-Ricci flow mainly based on \cite{BG,FOW,GKN}.

By a contact manifold it is understood a pair $(M,\eta)$ where $M$ is a smooth manifold of odd 
dimensions $(2n+1)$ together with $1$-form $\eta$ such that
\[
\eta \wedge (d\eta)^n \neq 0\,,
\]
is a volume form.

Associated with $\eta$ there is a unique vector field $\xi$, called Reeb vector field, characterized by
\[
\eta(\xi) = 1\quad \mbox{and} \quad d\eta (\xi,\cdot) = 0 \,.
\]

The Reeb vector field $\xi$ is a generator of $\mbox{ker} d\eta$ and there is a natural splitting 
of the tangent bundle of $M$
\[
TM = \mathcal{D} \oplus L_\xi \,, 
\]
where $L_\xi$ is a vertical subspace generated by $\xi$ and $\mathcal{D}$ is a horizontal distribution 
induced by $\mathcal{D} = \mbox{ker}\eta$.

A Sasakian manifold is a Riemannian manifold $(M, g)$ with the property that its metric
cone $(C(M),\bar{g})$
\[
C(M) = \mathbb{R}_{>0} \times M\,,\, \bar{g} = dr^2 + r^2 g\,,
\]
is K\"{a}hler. Here, $r\in (0,\infty)$ is the standard coordinate on the positive real line 
$\mathbb{R}_{>0}$.

On a contact manifold there is a tensor $\Phi$ of type $(1,1)$ satisfying 
\[
\Phi^2 = -\BOne + \eta \otimes \xi \quad \mbox{ and} \quad g(\Phi(X), \Phi(Y)) = 
g(X,Y) - \eta(X) \eta(Y) \,,
\]
for any vector fields $X,Y$ on $M$.

The restriction of the Sasaki metric $g$ to $\mathcal{D}$ gives a well-defined Hermitian metric
$g^T$. This Hermitian structure is in fact K\"{a}hler.

One can introduce local coordinates $(x, z^1, \dots ,z^n)$ on a small neighborhood of  $U=
I\times V$ with $I\in \mathbb{R}$ and $V \in \mathbb{C}^n$. In the chart $U$
we may write \cite{GKN}
\[
\xi =\frac{\partial}{\partial x}\,, 
\]
\[
\eta=dx+i\sum_{j=1}^n(K_{,j}{\rm d}z^j)-i\sum_{\bar{j}=1}^n(K_{,\bar{j}}
d\bar{z}^{j})\,,
\]
\[
d\eta=-2i\sum_{j,\bar{k}=1}^n K_{,j\bar{k}}
dz^{j}\wedge d\bar{z}^{k}\,,
\]
\[
g=\eta \otimes \eta + g^T =\eta \otimes \eta +
2\sum_{j,\bar{k}=1}^nK_{,j\bar{k}}dz^j{\rm d}\bar{z}^{k}\,,
\]
\[
\Phi=-i\sum_{j=1}^n[(\partial_j-iK_{,j}\partial_x)\otimes dz^j]+i\sum_{\bar{j}=1}^n
(\partial_{\bar{j}}+iK_{,\bar{j}}\partial_x)\otimes d{\bar{z}}^{j}]\,,
\]
where $K_{, j} = \frac{\partial}{\partial z^j}K$ and $K_{,j\bar{k}} = 
\frac{\partial^2}{\partial z^j \partial \bar{z}^k} K$. The function $K:U \rightarrow \mathbb{R}$ 
is the Sasakian analogue of the K\"{a}hler potential which does not depend on $x$, i.e. $\partial_x K=0$.

In what follows we consider deformations of the Sasaki structures
which preserve the Reeb vector field $\xi$. For this purpose it is necessary to introduce the 
basic forms. A $r$-form $\alpha$ on $M$ is called \emph{basic} if
\[
\iota_\xi \alpha =0\,,\quad \mathcal{L}_\xi \, \alpha =0\,,
\]
where $\mathcal{L}_\xi$ is the Lie derivative with respect to the vector field $\xi$.
In the system of coordinates $(x,z^1,\ldots,z^n)$ considered above, a basic $r$-form of type
$(p,q)\,,\, r=p+q$, has the form
\[
\alpha = \alpha_{i_1\cdots i_p \bar{j}_1 \cdots \bar{j}_q} dz^{i_1} \wedge
\cdots \wedge dz^{i_p} \wedge d\bar{z}^{j_1} \wedge \cdots \wedge d\bar{z}^{j_q}\,,
\]
where $\alpha_{i_1\cdots i_p \bar{j}_1 \cdots \bar{j}_q}$ does not depend on $x$. 
In particular a function $\varphi$ is basic if and only if $\xi(\varphi) = 0$. That is
the case of the Sasaki potential $K$.

Let $\varphi$ be a basic function and consider the deformation of the contact form $\eta$:
\begin{equation}\label{etadef}
\tilde{\eta}=\eta +d_{B}^{c}\varphi \,,
\end{equation}
where $d_{B}^{c}=\frac{i}{2}(\bar{\partial}_B-\partial_B)$ with
\[
\partial_B=\sum _{j=1}^{n}dz^j \frac{\partial}{\partial z^j}\,, \quad
\bar{\partial}_B=\sum _{j=1}^{n} d\bar{z}^{j} \frac{\partial}{\partial \bar{z}^{j}} \,.
\]

This deformation implies that other fundamental tensors are also modified:
\[
\begin{array}{rcl}
\tilde{\Phi} & = & \Phi - (\xi \otimes (d_{B}^{c}\varphi)) \circ \Phi \, ,\\
\tilde{g} & =  & d\tilde{\eta}\circ (\BOne \otimes \tilde{\Phi} ) +
\tilde{\eta}\otimes \tilde{\eta} \, ,\\
d\tilde{\eta}& = & d\eta+d_{B}d_{B}^{c}\varphi\,.
\end{array}
\]

To introduce the transverse K\"{a}hler-Ricci flow, also called Sasaki-Ricci flow, we
consider the flow  $(\xi, \eta(t), \Phi(t), g(t))$ with initial data
$(\xi, \eta(0), \Phi(0), g(0)) = (\xi, \eta, \Phi, g)$ generated by a basic function 
$\varphi(t)$. The Sasaki-Ricci flow equation is \cite{FOW}
\[
\frac{\partial g^T}{\partial t}=-Ric^T_{g(t)}+(2n+2)g^T(t)\,,
\]
where $Ric^T$  is the transverse Ricci curvature.
In the case of the deformation \eqref{etadef} with a basic function $\varphi$, in local coordinates 
the Sasaki-Ricci flow can be expressed as a parabolic Monge-Amp\`{e}re equation
\begin{equation}\label{Ricciflow}
\frac{\partial \varphi}{\partial t} = \ln \mbox{det}(g^T_{j\bar{k}} + \varphi_{j\bar{k}})
-\ln (\mbox{det} g^T_{j\bar{k}}) +  (2n+2) \varphi \,.
\end{equation}

\section{Sasaki-Einstein space $Y^{p,q}$ and Sasaki-Ricci \\ flow equation}

The metric of the Sasaki-Einstein space $Y^{p,q}$ is given by the line element \cite{MS}
\[
\begin{split}
ds^2 & = \frac{1- y}{6}( d \theta^2 + \sin^2 \theta\, d \phi^2)
+  \frac{1}{w(y)q(y)} dy^2
+ \frac{w(y)q(y)}{36} ( d\beta - \cos \theta \, d \phi)^2\\
& \quad  + \frac19
\left[ d\psi + \cos\theta \, d\phi + y(d\beta - \cos\theta \, d\phi)\right]^2
\,,
\end{split}
\]
where
\[
\begin{split}
w(y) & = \frac{2(a-y^2)}{1-y}\,,\\
q(y) & = \frac{a-3y^2 + 2 y^3}{a-y^2}\,,\\
f(y) & = \frac{a-2y + y^2}{6(a-y^2)}\,.
\end{split}
\]
Note that we have taken $\phi \rightarrow -\phi$ with respect to \cite{MS} and
consequently there are some differences of sign.

In the case of the space $Y^{p,q}$ the contact $1$-form $\eta$ is \cite{MS}
\[
\eta = \frac13 d\psi + \frac13 y \, d\beta  + \frac{1-y}3 \cos\theta \,d\phi\,,
\]
and the Reeb vector field is 
\begin{equation}\label{ReebY}
K_{\eta} = 3 \frac {\partial}{\partial \psi}\,.
\end{equation}

A detailed analysis  of the metric $Y^{p,q}$ \cite{GMSW} 
showed that 
it is globally well-defined and there are a countable infinite number of 
Sasaki-Einstein manifolds characterized by two relatively prime positive integers
$p, q$ with $p<q$. If $0< a<1$ the cubic equation
\begin{equation}\label{Q(y)}
Q(y) = a -3y^2 +2y^3 = \frac{1-y}2 w(y)\,q(y) = 0\,,
\end{equation}
has three real roots, one negative ($y_1$) and two positive, the smallest being $y_2$.
The coordinate $y$ ranges between the two smaller roots of the cubic equation \eqref{Q(y)},
i.e. $y_1\leq y \leq y_2$.

The angular coordinates span the ranges $ 0\leq \theta \leq \pi$, $ 0\leq \phi  \leq 2\pi$, 
$ 0\leq \psi \leq 2\pi$. In order to specify the range of the variable $\beta$, we note that 
it is connected with another variable $\alpha$
\[
 \beta = - (6\alpha + \psi)\,.
\]
The range of $\alpha$ is \cite{MS,GMSW}
\[
0 \leq \alpha \leq 2 \pi \ell\,,
\]
where
\[
\ell = \frac{q}{ 3q^2 - 2 p^2 + p(4 p^2 - 3 q^2 )^{1/2}}\,.
\]
The Reeb Killing vector field \eqref{ReebY} has compact orbits when $\ell$ is a rational number
and the corresponding $Y^{p,q}$ manifold is called quasi-regular. If $\ell$ is irrational
the orbits of the Reeb vector field do not close densely filling
the orbits of a torus and the Sasaki-Einstein manifold is said to be irregular.

For what follows it is useful to evaluate the following integrals
\[
f_1(y) = \exp\left( \int \frac1{H^2(y)} dy \right)
= \sqrt{(y-y_1)^{-\frac1{y_1}}(y_2-y)^{-\frac1{y_2}}(y_3-y)^{-\frac1{y_3}}}\,,
\]
\[
f_2(y) =\exp\left( \int \frac{y}{H^2(y)} dy \right)
=\frac1{\sqrt{Q(y)}} \,,
\]
where
\[
H^2(y) = \frac16 w(y)\,q(y) = \frac13 \frac{Q(y)}{1-y}\,.
\]

We introduce a local set of transverse complex coordinates \cite{MS,BLMP,CEPRV} addressing the 
transverse K\"{a}hler structure of $Y^{p,q}$:
\begin{equation}\label{cc}
\begin{split}
z^1 &= \tan\frac{\theta}2 e^{i\phi} \,,\\
z^2 &= \frac{\sin\theta}{f_1(y)} e^{i\beta}\,.
\end{split}
\end{equation}
These complex coordinates are not globally well-defined. This problem is discussed in 
\cite{BHOP} and a set of holomorphic coordinates on $Y^{p,q}$ was constructed.

We then get:
\begin{subequations}
\begin{align}
dz^1 &= \left( \frac12 \frac1{\cos^2\frac{\theta}2}d\theta
+ i\tan\frac{\theta}2 d\phi\right) e^{i\phi} \nonumber \,,\\
dz^2 &= \left( -\frac{f'_1}{f_1} dy + \cot\theta d\theta +i \,d\beta
\right)\frac{\sin\theta}{f_1(y)} e^{i\beta}\nonumber\\
&=\left( -\frac6{w(y)q(y)} dy + \cot\theta d\theta +i \,d\beta
\right)\frac{\sin\theta}{f_1(y)} e^{i\beta} \nonumber\,.
\end{align}
\end{subequations}

In terms of the complex coordinates \eqref{cc} the Sasaki-K\"{a}hler potential is
\[
K = \frac13 \left[ \left( 1 + \frac1{z^1 \bar{z}^1}\right) f_2(y)\right]
+ \frac16 \ln(z^1 \bar{z}^1)\,.
\]
Note that the additional term restores the correct form of the contact form $\eta$ of the space 
$Y^{p,q}$ without altering the transverse part of the metric.

We derive the local expressions of the derivatives of the Sasaki-K\"{a}hler potential. 
We can simplify calculation by defining \cite{BLMP}
\[
f^2_1(y) = \sigma = \frac{\sin^2\theta}{z^2 \bar{z}^2 }\,.
\]
First we evaluate

\[
\frac{d\,y}{d\sigma} = \frac{w(y)q(y)}{12 f_1^2(y)}\,,
\]
\[
\frac{\partial \sigma}{\partial z^1} =  \frac{\sigma}{z^1} \cos\theta\,, 
\]
\[
\frac{\partial \sigma}{\partial z^2} = - \frac{\sigma}{z^2}\,, 
\]
\[
\frac{\partial y}{\partial z^1}= \frac{w(y)q(y)}{12 z^1} \cos\theta\,, 
\]
\[
\frac{\partial y}{\partial z^2}= -\frac{w(y)q(y)}{12 z^2} \,, 
\]
\[
\frac{\partial}{\partial z^1} \left[ \ln f_2(y)\right]= \frac{y}{2z^1} \cos\theta\,,
\]
\[
\frac{\partial}{\partial z^2} \left[ \ln f_2(y)\right]= -\frac{y}{2z^2} \,.
\]

Derivatives of the Sasaki-K\"{a}hler potential are
\[
K_{,1} = \left[-\frac13\cos^2\frac{\theta}2
+\frac{y}6 \cos\theta \right]\cot\frac{\theta}2 e^{-i\phi}\,,
\]
\[
K_{,_2} = - \frac16 \frac{y f_1(y)}{\sin\theta} e^{-i\beta}\,,
\]
\[
K_{,1\bar{1}}= \frac13 (1-y) \cos^4\frac{\theta}2 + \frac{w(y)q(y)}{72}
\frac{\cos^2 \theta}{\tan^2\frac{\theta}2}\,,
\]
\[
K_{,22}= \frac{w(y)q(y)}{72} \frac{f^2_1(y)}{\sin^2\theta}\,,
\]
\[
K_{,1\bar{2}} =-\frac1{72} w(y)q(y) \cos\theta \cot\frac{\theta}2\, \frac{f_1(y)}{\sin\theta}
e^{-i\phi + i\beta}\,,
\]
\[
K_{,2\bar{1}} =-\frac1{72} w(y)q(y) \cos\theta \cot\frac{\theta}2\, \frac{f_1(y)}{\sin\theta}
e^{+i\phi - i\beta}\,.
\]

Consequently we get
\begin{equation}\label{gT}
\begin{split}
g^T &= 2\left( K_{,1\bar{1}} d z^1 d \bar{z}^1 +   K_{,1\bar{2}} d z^1 d \bar{z}^2
+ K_{,2\bar{1}} d z^2 d \bar{z}^1 + K_{,2\bar{2}} d z^2 d \bar{z}^2 \right)\\
&=\frac{1-y}6 \left( d\theta^2 + \sin^2\theta d\phi^2\right)+ \frac1{w(y)q(y)} dy^2 + 
\frac{w(y)q(y)}{36} \left(d\beta - \cos\theta d\phi \right)^2\,,
\end{split}
\end{equation}
\[
\mbox{det}\, g^T = 4(K_{,1\bar{1}}K_{,2\bar{2}} - K_{,1\bar{2}}K_{,2\bar{1}})
= \frac{w(y)q(y)}{216}f^2_1(y) (1-y) \cot^2\frac{\theta}2\,.
\]

For the purpose of studying the Sasaki-Ricci flow equation we evaluate the derivatives 
$\frac{\partial}{\partial z^j}$. We obtain for the derivatives involving the complex coordinate $z^1$:
\[
\frac{\partial}{\partial z^1}= \cos^2\frac{\theta}2 e^{-i\phi}
\left(\frac{\partial}{\partial \theta} -i \frac1{\sin\theta} \frac{\partial}{\partial \phi} \right) \,,
\]
\[
\frac{\partial}{\partial \bar{z}^1}= \cos^2\frac{\theta}2 e^{i\phi}
\left(\frac{\partial}{\partial \theta} +i \frac1{\sin\theta} \frac{\partial}{\partial \phi} \right) \,,
\]
\[
\frac{\partial^2}{\partial z^1 \partial \bar{z}^1}= \cos^4\frac{\theta}2 
\left(\frac{\partial^2}{\partial \theta^2} + \frac1{\sin^2\theta} \frac{\partial^2}{\partial \phi^2}
+ \cot\theta \frac{\partial}{\partial \theta}\right)\,.
\]

To have similar expressions for the partial derivatives involving $z^2$ we write
\[
z^2 = \rho e^{i\beta}\,,
\]
with
\[
\rho = \frac{\sin\theta}{f_1(y)}\,.   
\]
Then we get 
\[
\frac{\partial}{\partial z^2} = \frac{e^{-i\beta}}2
\left(\frac{\partial}{\partial \rho} - \frac{i}{\rho}\frac{\partial}{\partial \beta} \right)\,,
\]
\[
\frac{\partial}{\partial \bar{z}^2} = \frac{e^{i\beta}}2
\left(\frac{\partial}{\partial \rho} + \frac{i}{\rho}\frac{\partial}{\partial \beta} \right)\,,
\]
\[
\frac{\partial^2}{\partial z^2 \partial \bar{z}^2}= \frac14
\left(\frac{\partial^2}{\partial \rho^2} + \frac1{\rho^2} \frac{\partial^2}{\partial \beta^2}
+ \frac1\rho \frac{\partial}{\partial \rho}\right)\,,
\]
\[
\frac{\partial^2}{\partial z^1 \partial \bar{z}^2}= \frac{\cos^2\frac{\theta}2}2 e^{i\beta-i\phi}
\left(\frac{\partial^2}{\partial \rho \partial\theta} + \frac{i}{\rho} 
\frac{\partial^2}{\partial \beta \partial\theta} - \frac{i}{\sin\theta}
\frac{\partial^2}{\partial \rho \partial\phi} + \frac1{\rho\sin\theta} 
\frac{\partial^2}{\partial \beta \partial\phi}\right)\,,
\]
\[
\frac{\partial^2}{\partial z^2 \partial \bar{z}^1}= \frac{\cos^2\frac{\theta}2}2 e^{-i\beta+i\phi}
\left(\frac{\partial^2}{\partial \rho \partial\theta} - \frac{i}{\rho} 
\frac{\partial^2}{\partial \beta \partial\theta} + \frac{i}{\sin\theta}
\frac{\partial^2}{\partial \rho \partial\phi} + \frac1{\rho\sin\theta} 
\frac{\partial^2}{\partial \beta \partial\phi}\right)\,.
\]

Sasaki-Ricci flow equation \eqref{Ricciflow} on $Y^{p,q}$ becomes:
\begin{equation}\label{RF2}
\begin{split}
\frac{d\varphi}{dt} = &\ln\Biggl\{ \varphi_{,1\bar{1}}\varphi_{,2\bar{2}}-\varphi_{,1\bar{2}}\varphi_{,2\bar{1}}
+\left[\frac13 (1-y) \cos^4\frac{\theta}2 + \frac{w(y)q(y)}{72}
\frac{\cos^2 \theta}{\tan^2\frac{\theta}2}\right] \varphi_{,2\bar{2}}\\
& +\frac{w(y)q(y)}{72} \frac{f^2_1(y)}{\sin^2\theta}\varphi_{,1\bar{1}} 
+\frac{w(y)q(y)}{72} \cos\theta \cot\frac{\theta}2\, \frac{f_1(y)}{\sin\theta}
e^{-i\phi + i\beta}\varphi_{,2\bar{1}}\\
& +\frac{w(y)q(y)}{72} \cos\theta \cot\frac{\theta}2\, \frac{f_1(y)}{\sin\theta}
e^{i\phi - i\beta}\varphi_{,1\bar{2}} 
+ \frac{w(y)q(y)}{216}f^2_1(y) (1-y) \cot^2\frac{\theta}2 \Biggr\}\\
&-\ln \left[\frac{w(y)q(y)}{216}f^2_1(y) (1-y) \cot^2\frac{\theta}2\right] +6 \varphi\,.
\end{split}
\end{equation}

Let us assume that the dependence of $\varphi$ on $z^1$ and $z^2$ separates
\[
\varphi = f(t) \left[ g_1(z^1,\bar{z}^1) + g_2(z^2,\bar{z}^2) \right]\,,
\]
where the functions $\varphi, g_1, g_2$ are to be determined.
In this case the mixed derivatives $\varphi_{,1\bar{2}}$ and $\varphi_{,2\bar{1}}$ vanish
and the evaluation of the derivatives $\varphi_{,1\bar{1}}, \varphi_{,2\bar{2}}$ becomes simpler:
\begin{equation}\label{11}
\varphi_{,1\bar{1}}= \cos^4\frac{\theta}2 
\left(\frac{\partial^2 g_1}{\partial \theta^2}
+ \frac1{\sin^2\theta} \frac{\partial^2 g_1}{\partial \phi^2}
+ \cot\theta \frac{\partial g_1}{\partial \theta}\right) f(t)\,,
\end{equation}
\begin{equation}\label{22}
\varphi_{,2\bar{2}}= \frac14
\left(\frac{\partial^2 g_2}{\partial \rho^2} 
+ \frac1{\rho^2} \frac{\partial^2 g_2}{\partial \beta^2}
+ \frac1\rho \frac{\partial g_2}{\partial \rho}\right)f(t)\,.
\end{equation}

Explicit solutions of the Sasaki-Ricci flow equation can be obtained assuming
\begin{equation}\label{11c}
\varphi_{,1\bar{1}}= \cos^4\frac{\theta}2  c_1 f(t)\,,
\end{equation}
\begin{equation}\label{22c}
\varphi_{,2\bar{2}}= c_2 f(t)\,,
\end{equation}
where $c_1, c_2$ are arbitrary constants.

Analogous assumptions were implied in the study of the Sasaki-Ricci flow on $T^{1,1}$ 
\cite{SVV19}. Using \eqref{11c} from \eqref{11} we obtain for $g_1$ an explicit expression:
\[
g_1 (\theta,\phi)= \frac{d_1}2 \phi^2 + e_1 \ln \tan \frac{\theta}2 - 
\frac{d_1}{2} \left(\ln \tan \frac{\theta}2\right)^2 - 
c_1 \ln \sin\theta  \,,
\]
involving the arbitrary constants $c_1, d_1, e_1$.

Similarly, assuming \eqref{22c} equation \eqref{22} has a simple solution:
\[
g_2 (\rho,\beta) = \frac{d_2}2 \beta^2 + c_2 \rho^2 + e_2 \ln \rho - 
\frac{d_2}2 (\ln \rho)^2 \,,
\]
where $c_2, d_2, e_2$ are other arbitrary constants.

Let us remark that for $c_1 = c_2 =0$, the transverse metric remains unaltered 
and the Sasaki-Ricci flow equation reduces to
\[
\frac{d\varphi}{dt} =6 \varphi \,,
\]
with the obvious solution taking the initial condition $f(0) = 0$:
\[
f(t) = e^{6t} -1\,. 
\]

The deformed contact form can be evaluated according to \eqref{etadef}:
\[
\tilde{\eta} = \eta +  \frac{i}2 \sum_m \varphi_{,m} d z^m -
\frac{i}2 \sum_m \varphi_{,\bar{m}} d \bar{z}^m
\]

To summarize the above analysis we have the following outcome:
\begin{proposition}
The families of basic functions
\[
\begin{split}
\varphi (t) = (e^{6t} -1)&\left[ \frac{d_1}2 \phi^2 + e_1 \ln \tan \frac{\theta}2 - 
\frac{d_1}2 \left(\ln \tan \frac{\theta}2 \right)^2 \right.\\
&\left. +\frac{d_2}2 \beta^2 + e_2 \ln \rho - \frac{d_2}2 (\ln \rho)^2 \right] \,,
\end{split}
\]
with $d_j, e_j$ arbitrary constants,
stand as solutions of the transverse K\"{a}hler-Ricci flow equation on the manifold $Y^{p,q}$.

The corresponding deformed contact structures remain Sasaki-Einstein with the contact forms
\[
\begin{split}
\tilde{\eta} = \eta + \frac{e^{6t} -1}2 &\left[ \frac{d_1 \phi}{\sin \theta} d\theta + \left( -e_1 + 
d_1 \ln \tan \frac{\theta}2 \right) d\phi \right. \\
&\left. + \frac{d_2 \beta}{\rho} \, d\rho + (- e_2 + d_2 \ln \rho ) d\beta 
\right] \,.
\end{split}
\]
\end{proposition}

When $c_j \neq 0$, Sasaki-Ricci flow equation \eqref{RF2} becomes more involved. In spite of the 
fact that we have not an explicit solution for the basic function $\varphi (t)$, the Ricci flow produces
new Sasaki structures with deformations of the transverse  metric $g^T$ \eqref{gT} of the source $Y^{p,q}$ 
metric.
\begin{proposition}
The deformed contact structures with the contact forms
\[
\tilde{\eta} = \eta + \frac{f(t)}2 \left[c_1 \cos \theta\, d\phi  - c_2 \rho^2 d \beta\right] \,,
\] 
with $c_j$ arbitrary constants, remain Sasaki with the deformed metrics
\[
\begin{split}
\tilde{g} & = \tilde{\eta} \otimes \tilde{\eta} +  g^T + \sum_{j=1}^{2} \phi_{j\tilde{j}} dz^j d\tilde{z}^j\\
& = \tilde{\eta} \otimes \tilde{\eta} +  g^T + f(t) \left[ \frac{c_1}4 (d \theta^2 + \sin^2 \theta \,d \phi^2) +
c_2(d \rho^2 + \rho^2 d \beta^2)\right] \,.
\end{split}
\]
\end{proposition}

\section{Concluding remarks}

To deform the Sasakian structures we exploit the transverse structure of Sasaki manifolds. The Sasaki-Ricci
flow is a transverse K\"{a}hler-Ricci flow which deforms the transverse K\"{a}hler structure.

To perturb the Sasakian structure, we keep the Reeb field fixed and let vary the contact form $\eta$ by
modifying it with a basic function as in \eqref{etadef}. For small perturbing basic functions, the Sasakian
structure of the manifold is preserved.

Starting with a Sasaki-Einstein manifold $Y^{p,q}$, as a seed, we generate families of new Sasakian structures. 
We are able to find explicit solutions of the Sasaki-Ricci flow equation depending on some
arbitrary constants.

Recently we discussed the complete integrability on $T^{1,1}$ and $Y^{p,q}$ spaces
\cite{MV12,BV,MV17,SVV15}.
It would be interesting to investigate the contact Hamiltonian systems \cite{BCT,LV}, complete integrability 
and action angle variables on the perturbed Sasaki-Einstein spaces $Y^{p,q}$.

It is worth extending the study of Sasaki-Ricci flow on higher dimensional Sasaki-Einstein spaces as well 
as other contact structures as $3$-Sasaki structures \cite{BG99} or mixed $3$-structures \cite{IVV}.

\section*{Acknowledgments}

This work has been supported by the project {\it NUCLEU PN 19 06 01 
01/2019}.

\end{document}